\documentclass[aps,pra,twocolumn,superscriptaddress]{revtex4-1}
\usepackage[dvipdfmx]{graphicx,epsfig,hyperref}
\usepackage{amsmath,amssymb,amsfonts,physics}
\usepackage{bm,braket,comment,color}
\newcommand*{\beq}{\begin{equation}}
\newcommand*{\eeq}{\end{equation}}
\newcommand*{\beqa}{\begin{eqnarray}}
\newcommand*{\eeqa}{\end{eqnarray}}
\newcommand*{\bseq}{\begin{subequations}}
\newcommand*{\eseq}{\end{subequations}}
\newcommand*{\bal}{\begin{aligned}[b]}
\newcommand*{\eal}{\end{aligned}}
\newcommand*{\bpm}{\begin{pmatrix}}
\newcommand*{\epm}{\end{pmatrix}}
\newcommand*{\rc}{\mathrm{RC}}
\newcommand*{\J}{\mathrm{J}}
\begin{document}

\title{Quantum and thermal fluctuations in the dynamics of a resistively and capacitively shunted Josephson junction}
\author{Koichiro Furutani}
\email{koichiro.furutani@phd.unipd.it}
\affiliation{Dipartimento di Fisica e Astronomia ``Galileo Galilei'', 
Universit\`a di Padova, via Marzolo 8, 35131 Padova, Italy}
\affiliation{Istituto Nazionale di Fisica Nucleare (INFN), Sezione di Padova, 
via Marzolo 8, 35131 Padova, Italy}
\author{Luca Salasnich}
\affiliation{Dipartimento di Fisica e Astronomia ``Galileo Galilei'', 
Universit\`a di Padova, via Marzolo 8, 35131 Padova, Italy}
\affiliation{Istituto Nazionale di Fisica Nucleare (INFN), Sezione di Padova, 
via Marzolo 8, 35131 Padova, Italy}
\affiliation{CNR-INO, via Nello Carrara, 1 - 50019 Sesto Fiorentino, Italy}

\date{today}
\begin{abstract}
We theoretically investigate the phase and voltage correlation dynamics, which includes both the deterministic contribution and stochastic fluctuations, under a current noise generated by a resistor including thermal and quantum fluctuations in a resistively and capacitively shunted Josephson junction. 
An external current is found to shift and intensify the deterministic contributions in phase and voltage. 
In addition to effects of external current, we observe the relaxation of autocorrelation functions of phase and voltage, which includes the variances due to the current noise, to finite values in the long-time limit. 
In particular, we find that the asymptotic correlations depend on the resistance as a consequence of quantum effects.
We also find an earlier decay of coherence at a higher temperature in which thermal fluctuations dominate over quantum ones. 
These theoretical predictions can be tested in the next future experiments.
\end{abstract}
\maketitle

\section{Introduction}

Noise is ubiquitous in diverse physical systems such as thermal noise in classical many-body systems and $1/f$ noise in electronic devices \cite{tauber,kamenev}. It is one of the inevitable effects in a realistic system, but it also plays an important role in physical properties \cite{johnson,nyquist,halperin,caldeira,dalla10,diehl,dalla12}. An origin of the noise is the thermal fluctuations, which is usually treated as a Gaussian white noise \cite{tauber,kamenev}. Another crucial origin of the noise at low temperature is the quantum fluctuations reflecting the quantum mechanical property of a system \cite{metiu,ford,brandt,beck,callen,ingold,koch80,koch82,levinson}. These kinds of noise make the physical observables time-dependent distinct from equilibrium cases.

\begin{figure}[t]
\centering
\includegraphics[width=70mm]{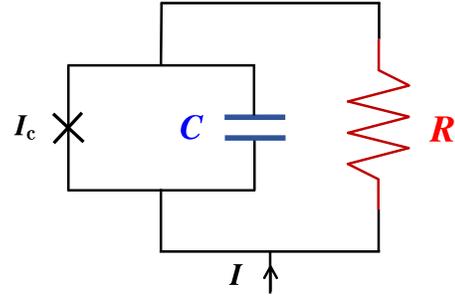}
\caption{RCSJ junction with resistance $R$ and capacitance $C$ that we study in this paper. Here $I_{\mathrm{c}}$ denotes the critical current in the Josephson junction and $I$ is the external current.}
\label{circuit}
\end{figure}

In particular, the noise spectrum in a resistively and capacitively shunted Josephson (RCSJ) junction has been extensively investigated both theoretically and experimentally \cite{koch80,koch82,brandt,levinson}. The schematic picture is given in Fig. \ref{circuit} with a resistor, a capacitor, and a Josephson junction in parallel. 
The RCSJ junction exhibits interesting physics such as Schmid-Bulgadaev transition, which states the presence of the transition to an insulating phase from the superconducting phase at a larger resistance than a critical resistance, and is an important platform to investigate quantum dynamics \cite{schmid,bulgadaev,ingold99,kimura,lukyanov,yagi,blais,murani,glazman}. 
The current noise in this RCSJ junction originates from the shunted resistor in a RCSJ junction \cite{koch80, koch82}. Reference \cite{koch82} has experimentally observed the current noise spectrum at low temperature. 
This measurement revealed the presence of a zero-point energy term in the current noise, which justifies our treatment of quantum noise in this paper. 
Based on the several discussions on the quantum noise spectrum with respect to the current or voltage, it is crucial to determine the dynamics of correlations in a RCSJ junction subject to the quantum current noise.

In this paper, we present the effects of noise including thermal and quantum fluctuations in a RCSJ junction on the dynamics of correlation functions, which includes the deterministic part and stochastic fluctuations due to the current noise induced by a resistor, in order to establish a systematic understanding of dynamics in a noisy Josephson junction. Starting from the equation of motion in terms of the phase in the presence of the external current as well as the current noise including thermal and quantum fluctuations, we analytically obtain the dynamics of relative phase and voltage within the linear regime. 
Then we observe the relaxation of autocorrelation functions, composed of the deterministic contributions and variances, with respect to the phase and voltage. 
We find that the asymptotic correlation in the long-time limit of phase, as well as that of voltage, involves dependence on the damping constant. Since in the high-temperature limit, the correlations are independent, the dependence on the damping constant indicates the emergence of quantum effects at low temperature. 
Similar dependence on damping has been observed in a Bose Josephson junction, which involves an intrinsic coupling between the Josephson mode and bath mode and exhibits different behaviors compared to our system of the RCSJ junction \cite{binanti}.
We expect that this quantum effect as a dependence of the correlations on the damping constant can be experimentally measured in the RCSJ junction.
With this respect, our work of dynamics in a RCSJ junction provides a benchmark to test quantum effects in a noisy system out of equilibrium.

\section{Equation of motion in a RCSJ junction}\label{SecII}

\begin{figure}[t]
\centering
\includegraphics[width=80mm]{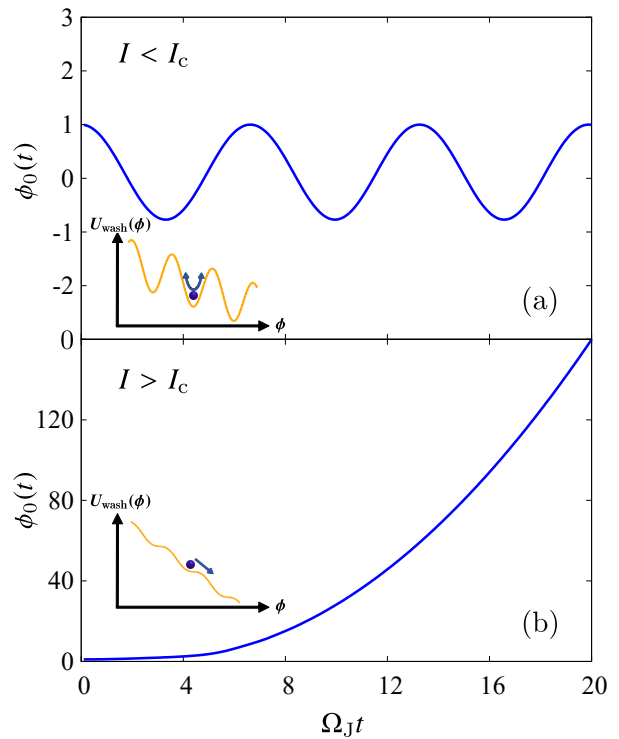}
\caption{Deterministic part of phase $\phi_{0}(t)$ subject to the washboard potential in Eq. \eqref{washboard} in the absence of friction. We set the initial condition $\phi_{0}(0)=1$ and $\dot{\phi}_{0}(0)=1$ for brevity. The upper panel (a) shows the case of $I<I_{\mathrm{c}}$, while the lower panel (b) displays the case of $I>I_{\mathrm{c}}$. In the former case, the phase oscillates around an extremum of the washboard potential. On the other hand, in the latter case, the potential has no extremum leading to the running state with respect to the phase.}
\label{phiwash}
\end{figure}

We consider a RCSJ junction composed of a resistor, a capacitor, and a Josephson junction in parallel as depicted in Fig. \ref{circuit}. 
It is described by
\beqa
&C\dot{V}(t)& +\frac{V(t)}{R}+\frac{\partial U_{\mathrm{wash}}(\phi)}{\partial\phi} = \eta(t), 
\label{VI} \\
&\dot{\phi}(t)& = \frac{2e V(t)}{\hbar},
\label{Vphi}
\eeqa
where $\phi(t)$, $V(t)$, $C$, and $R$ are respectively the superconducting phase, voltage, capacitance, and resistance. In Eq. \eqref{VI},
\beq
U_{\mathrm{wash}}(\phi)=-I_{\mathrm{c}}\cos{\phi(t)}-I\phi(t),
\label{washboard}
\eeq
is the tilted washboard potential with the critical current $I_{\mathrm{c}}$ \cite{blackburn, lukyanov}. 
For a small external current $I<I_{\mathrm{c}}$, the washboard potential has potential minima $\sin{\phi}=I/I_{\mathrm{c}}$ while if the external current exceeds $I_{\mathrm{c}}$, it has no potential minimum and it may drive the phase into a running state as shown in Fig. \ref{phiwash}. 
Throughout this paper, we consider the case with a small external current as in Fig. \ref{phiwash}(a) excluding the running state.
Equation \eqref{Vphi} provides the relation between the phase and the voltage. In Eq. \eqref{Vphi}, $e$ is the elementary charge and $\hbar$ is the reduced Planck constant. 
In this paper, we deal with the phase, voltage, and current noise as classical quantities. Then, the current noise $\eta(t)$ in Eq. \eqref{VI}, which originates from the shunted resistor, satisfies
\bseq
\beqa
&\left\langle\eta(t)\right\rangle =0, 
\\
&\bal
\displaystyle\int^{\infty}_{-\infty}dt \left\langle\eta(t)\eta(0)\right\rangle e^{-i\omega t}
&=\frac{2}{R}\hbar\omega\coth{\left(\frac{\hbar\omega}{2k_{\mathrm{B}}T}\right)} \\
&\equiv \Gamma(\omega),
\eal
\label{qnoiseb}
\eeqa
\label{qnoise}
\eseq
where $k_{\mathrm{B}}$ is the Boltzmann constant and $T$ is the temperature. The average $\langle\cdots\rangle$ stands for the Gaussian average with respect to the colored noise $\eta(t)$. In the classical limit $\hbar\omega \ll k_{\mathrm{B}}T$, Eq. \eqref{qnoiseb} reproduces the classical fluctuation-dissipation relation as given in Appendix \ref{AppA}. On the other hand, at $T=0$, Eq. \eqref{qnoiseb} results in $\Gamma(\omega)\to2\hbar\abs{\omega}/R$ \cite{caldeira,dalla10}. The $\omega$ dependence of the correlation in Eqs. \eqref{qnoise} indicates that the quantum noise $\eta$ is the colored noise and approaches the white one in the classical limit. This current noise spectrum of Eq. \eqref{qnoiseb} has been experimentally measured in Ref. \cite{koch82} and includes the zero-point fluctuations in the shunted resistor.

The equations given in Eq. \eqref{VI} and Eq. \eqref{Vphi} provide the equation of motion with respect to the relative phase $\phi$ as \cite{brandt}
\beq
\frac{\hbar C}{2e}\ddot{\phi}(t)+\frac{\hbar}{2eR}\dot{\phi}(t)+\frac{\partial U_{\mathrm{wash}}(\phi)}{\partial\phi}=\eta(t) .
\label{EoMeta}
\eeq
The solution of Eq. \eqref{EoMeta} can be written as \cite{brandt}
\beq
\phi(t)=\phi_{0}(t)+\delta\phi(t),
\eeq
where $\phi_{0}$ is the deterministic part of the relative phase and $\delta\phi$ represents the stochastic part due to the current noise. 
Assuming $\abs{\delta\phi(t)}\ll\abs{\phi_{0}(t)}$, each of the components satisfies \cite{brandt}
\beq
\frac{\hbar C}{2e}\ddot{\phi}_{0}(t)+\frac{\hbar}{2eR}\dot{\phi}_{0}(t)+\frac{\partial U_{\mathrm{wash}}(\phi_{0})}{\partial\phi_{0}}=0,
\label{EoMphi0}
\eeq
\beq
\frac{\hbar C}{2e}\delta\ddot{\phi}(t)+\frac{\hbar}{2eR}\delta\dot{\phi}(t)+I_{\mathrm{c}}\cos{\phi_{0}(t)}\delta\phi(t)=\eta(t).
\label{EoMdeltaphi}
\eeq
In Ref. \cite{brandt}, Brandt {\it et al.} have employed the approximation $\abs{\delta\phi(t)}\ll\abs{\phi_{0}(t)}$ under the assumption of the small current noise. We also adopt this assumption throughout this paper. 
Hereafter we use the following notations
\beq
\Omega_{\J}^{2}\equiv\frac{2e}{\hbar}\frac{I_{\mathrm{c}}}{C}=\frac{2\pi I_{\mathrm{c}}}{\Phi_{0}C}, 
\eeq
\beq
\Omega_{\rc}\equiv\frac{1}{2RC}, 
\eeq
where $\Phi_{0}=h/(2e)$ is the magnetic flux quantum, $\Omega_{\J}$ is the Josephson plasma frequency, and $\Omega_{\rc}$ is related to the resistance leading to a damping in phase dynamics; one can write Eq. \eqref{EoMphi0} and Eq. \eqref{EoMdeltaphi} as
\beq
\ddot{\phi}_{0}+2\Omega_{\rc}\dot{\phi}_{0}+\Omega_{\J}^{2}\sin{\phi_{0}}=\Omega_{\J}^{2}\frac{I}{I_{\mathrm{c}}},
\label{EoMphi0ad}
\eeq
\beq
\delta\ddot{\phi}+2\Omega_{\rc}\delta\dot{\phi}+\Omega_{\J}^{2}\delta\phi\cos{\phi_{0}}=\Omega_{\J}^{2}\frac{\eta}{I_{\mathrm{c}}}.
\label{EoMdeltaphiad}
\eeq

\subsection{Linear analysis in the absence of external current}\label{SecIIA}

In this paper, we focus on the linear regime in which the washboard potential in Eq. \eqref{washboard} can be well approximated to a harmonic potential in addition to the term that involves the small external current corresponding to the upper case in Fig. \ref{phiwash}(a). 
In the linear regime $\sin{\phi_{0}}\simeq\phi_{0}$ and $\cos{\phi_{0}}\simeq1$ in the absence of the external current $I=0$, the equations of motion in Eqs. \eqref{EoMphi0ad} and \eqref{EoMdeltaphiad} can be written as
\beq
\ddot{\phi}_{0}+2\Omega_{\rc}\dot{\phi}_{0}+\Omega_{\J}^{2}\phi_{0}=0,
\label{EoMphi0line}
\eeq
\beq
\delta\ddot{\phi}+2\Omega_{\rc}\delta\dot{\phi}+\Omega_{\J}^{2}\delta\phi=\frac{\Omega_{\J}^{2}}{I_{\mathrm{c}}}\eta.
\label{EoMdelphiline}
\eeq
We can obtain the root of this approximated equation of motion as
\beq
\phi(t)=\phi_{0}(t)+\int^{t}_{0}dt_{1}G(t-t_{1})\eta(t_{1}) ,
\eeq
and the two-point correlation function
\beq
\bal
\left\langle\phi(t)\phi(t')\right\rangle
&=\phi_{0}(t)\phi_{0}(t') \\
&+\int^{t}_{0}dt_{1}\int^{t'}_{0}dt_{2}G(t-t_{1})G(t'-t_{2})\left\langle\eta(t_{1})\eta(t_{2})\right\rangle,
\eal
\label{rootEoM}
\eeq
where $\phi_{0}(t)$ is the solution of Eq. \eqref{EoMphi0line}, and
\beq
G(t)=\frac{\Omega_{\J}^{2}}{I_{\mathrm{c}}}\frac{e^{-\Omega_{\rc}t}}{\sqrt{\Omega_{\J}^{2}-\Omega_{\rc}^{2}}}\sin{\left(\sqrt{\Omega_{\J}^{2}-\Omega_{\rc}^{2}}t\right)}\theta(t).
\label{Gt}
\eeq
 Using Eqs. \eqref{qnoise}, the second term in the right hand side of Eq. \eqref{rootEoM} can be written as
\beq
\bal
&\int^{t}_{0}dt_{1}\int^{t'}_{0}dt_{2}G(t-t_{1})G(t'-t_{2})\left\langle\eta(t_{1})\eta(t_{2})\right\rangle \\
&=\int^{\infty}_{-\infty}\frac{d\omega}{2\pi}\Gamma(\omega)G_{t}(\omega)G_{t'}(-\omega)e^{i\omega(t-t')} .
\eal
\eeq
where
\beq
G_{t}(\omega)\equiv\int^{t}_{0}dt_{1}G(t_{1})e^{-i\omega t_{1}} .
\label{Gomega}
\eeq
The explicit expression of Eq. \eqref{Gomega} is given in Appendix \ref{AppB}. Note that Eq. \eqref{rootEoM} is real, while Eq. \eqref{Gomega} is a complex function. The dynamics in the absence of the noise as a solution of Eq. \eqref{EoMphi0ad} is
\beq
\phi_{0}(t)=\frac{2\pi V_{0}}{\Phi_{0}}\frac{e^{-\Omega_{\rc}t}}{\sqrt{\Omega_{\J}^{2}-\Omega_{\rc}^{2}}}\sin{\left(\sqrt{\Omega_{\J}^{2}-\Omega_{\rc}^{2}}t\right)}.
\label{phi0I0}
\eeq
Hence we finally obtain the correlation function including the quantum noise as
\beq
\bal
&\left\langle\phi(t)\phi(t')\right\rangle
=\phi_{0}(t)\phi_{0}(t') \\
&+\frac{2}{R}\int^{\infty}_{-\infty}\frac{d\omega}{2\pi}\hbar\omega\coth{\left(\frac{\hbar\omega}{2k_{\mathrm{B}}T}\right)}G_{t}(\omega)G_{t'}(-\omega)e^{i\omega(t-t')}.
\eal
\label{phiphiq}
\eeq
The energy integral in Eq. \eqref{phiphiq} involves a logarithmic UV divergence due to the zero-point fluctuations in the noise spectrum \cite{koch82,koch80}. In our calculations below, we restrict the energy range as $-\Delta \le \hbar\omega \le \Delta$, where $\Delta$ is the energy gap of the superconductor. This gap is related to the critical current by the Ambegaokar-Baratoff formula as $I_{\mathrm{c}}=\pi \Delta/(2eR)\tanh{\left[\Delta/(2k_{\mathrm{B}}T)\right]}$ \cite{ambeg,sauls}.
One may think that, according to Eq. \eqref{phi0I0}, $\phi_{0}(t\to\infty)\to0$ leads to the breakdown of the approximation $\abs{\delta\phi(t)}\ll\abs{\phi_{0}(t)}$ that we assumed in Sec. \ref{SecII}. It is true that the higher order in $\delta\phi(t)$ can affect the correlations. 
In the long-time limit in which $\phi_{0}$ vanishes, however, the results under the approximation can be valid. Setting $\phi_{0}(t)=0$ in the absence of the external current, one obtains
\beq
\delta\ddot{\phi}+2\Omega_{\rc}\delta\dot{\phi}+\Omega_{\J}^{2}\sin{\delta\phi}=\Omega_{\J}^{2}\frac{\eta}{I_{\mathrm{c}}}.
\label{EoMdeltaphiwash}
\eeq
Hence, within the linear regime with respect to $\delta\phi$, we obtain totally the same equation as Eq. \eqref{EoMdelphiline} and our results would be valid even in the long-time limit.

\begin{figure}[t]
\centering
\includegraphics[width=80mm]{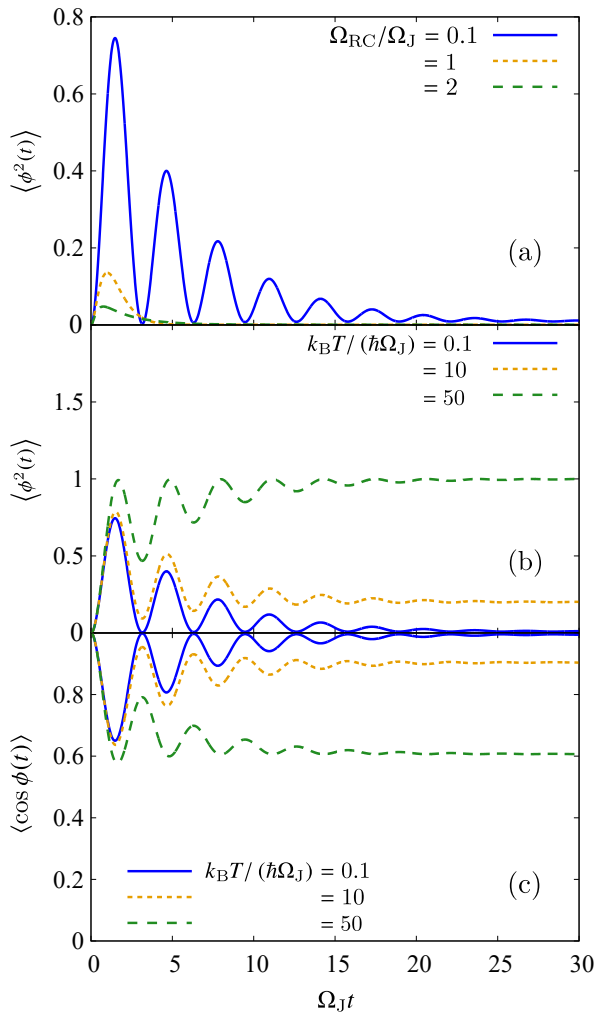}
\caption{Time evolution of the autocorrelation function with respect to the phase $\phi(t)$. 
The upper panel (a) displays the results for $k_{\mathrm{B}}T/\left(\hbar\Omega_{\J}\right)=0.1$. The blue solid, orange dotted, and green dashed lines respectively stand for the results for $\Omega_{\rc}/\Omega_{\J}=0.1$ (underdamped), $\Omega_{\rc}/\Omega_{\J}=1$ (critically damped), and $\Omega_{\rc}/\Omega_{\J}=2$ (overdamped). 
The panel (b) displays the results for $\Omega_{\rc}/\Omega_{\J}=0.1$. The blue solid, orange dotted, and green dashed lines respectively stand for the results for $k_{\mathrm{B}}T/\left(\hbar\Omega_{\J}\right)=0.1, 10, 50$. 
The lower panel (c) displays the coherence factor for $\Omega_{\rc}/\Omega_{\J}=0.1$.
We set $2e\Omega_{\J}/I_{\mathrm{c}}=10^{-2}$ and $V_{0}=\hbar\Omega_{\J}/(2e)$.}
\label{phiphi}
\end{figure}

Figure \ref{phiphi}(a) shows the dynamics of the autocorrelation function of the relative phase $\langle\phi^{2}(t)\rangle$ for $k_{\mathrm{B}}T=0.1\hbar\Omega_{\J}$ with different damping. In the following, we set $2e\Omega_{\J}/I_{\mathrm{c}}=10^{-2}$ and $V_{0}=\hbar\Omega_{\J}/(2e)$ for brevity. 
Experimentally, in Ref. \cite{devoret} for instance, they used $I_{\mathrm{c}}=9.489$ $\mathrm{\mu A}$ and $\Omega_{\J}=67.4$ $\mathrm{GHz}$, which corresponds to $2e\Omega_{\J}/I_{\mathrm{c}}\simeq 2.3\times 10^{-3}$ and $V_{0}\simeq 2.2\times10^{-5}$ $\mathrm{V}$.
Based on these experimental values, we chose the fixed parameter $2e\Omega_{\J}/I_{\mathrm{c}}=10^{-2}$. 
One can see that the correlation is suppressed as one increases the damping coefficient $\Omega_{\rc}/\Omega_{\J}$. 
This is a quite intuitive behavior because the large damping constant leads to an earlier exponential decay of the phase correlation according to Eq. \eqref{Gt}. 
In addition, the energy gap $\Delta\sim \Omega_{\J}/\Omega_{\rc}$ at low temperatures as an energy cutoff in Eq. \eqref{phiphiq} is also responsible for this strong suppression with a smaller resistance.
In the experiment in Ref. \cite{devoret}, Devoret {\it et al.} measured $R=190$ $\mathrm{\Omega}$ and $C=6.35$ $\mathrm{pF}$ resulting in $\Omega_{\rc}/\Omega_{\J}\simeq6.2\times10^{-3}$, which corresponds to the case with a tiny damping constant. 
Only in the long-time limit $t\to\infty$ does the energy integral in Eq. \eqref{phiphiq} converge without any cutoff energy.
Remarkably, we found that the asymptotic correlation still depends on the damping constant as a consequence of quantum fluctuations in the current noise [see Eq. \eqref{phiphiqa} in Appendix \ref{AppC}]. In the classical limit, as in Eq. \eqref{phiphicla}, it is independent of the damping. 
This quantum effect as the dependence on the resistance of the correlations in the long-time limit can be measured experimentally.
In the underdamped limit $\Omega_{\rc} \ll \Omega_{\J}$, in particular, it reduces to $2e\Omega_{\J}/I_{\mathrm{c}}\cdot\coth{\left[\hbar\Omega_{\J}/\left(2k_{\mathrm{B}}T\right)\right]}$, which recovers the classical asymptotic value $2k_{\mathrm{B}}T/\left(\hbar\Omega_{\J}\right)$ in the classical limit $k_{\mathrm{B}}T\gg\hbar\Omega_{\J}$ (see Appendix \ref{AppC}).

\begin{figure}[t]
\centering
\includegraphics[width=80mm]{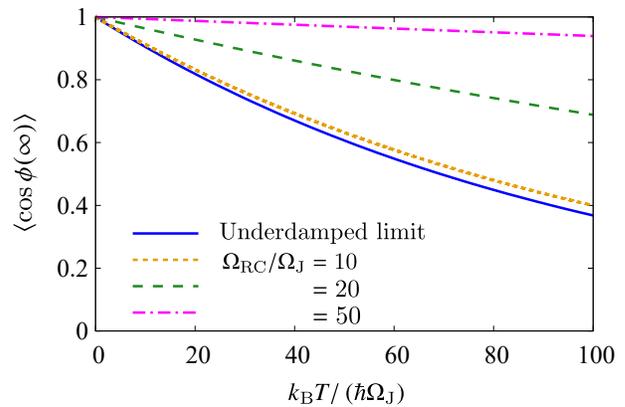}
\caption{Temperature dependence of the coherence factor in the long-time limit $\left\langle\cos{\phi(t\to\infty)}\right\rangle$.
The solid line stands for the result in the underdamped limit $\Omega_{\rc}/\Omega_{\J}\to0$ [see Eq. \eqref{phiphiua}]. The dotted, dashed, and dotted-dashed lines represent the results for $\Omega_{\rc}/\Omega_{\J}=10, 20, 50$, respectively, obtained by Eq. \eqref{phiphiqa}.  We set $2e\Omega_{\J}/I_{\mathrm{c}}=10^{-2}$ and $V_{0}=\hbar\Omega_{\J}/(2e)$.}
\label{coherenceasym}
\end{figure}

Figure \ref{phiphi}(b) illustrates the numerical results with different temperature in the underdamped regime $\Omega_{\rc}=0.1\Omega_{\J}$. 
It shows that the autocorrelation is enhanced in the high-temperature region due to the dominant thermal fluctuations compared to the low-temperature region in which quantum fluctuations dominate over thermal ones. 
It also indicates that the asymptotic value gets closer to the classical one $2k_{\mathrm{B}}T/\left(\hbar\Omega_{\J}\right)$ as one increases temperature, as expected.

Figure \ref{phiphi}(c) displays the time evolution of the coherence factor for $\Omega_{\rc}/\Omega_{\J}=0.1$. 
Using the Gaussian property of the noise $\eta$, one can compute it by
\beq
\left\langle\cos{\phi(t)}\right\rangle=\cos{\phi_{0}(t)}e^{-\frac{1}{2}\left[\left\langle\phi^{2}(t)\right\rangle-\phi_{0}^{2}(t)\right]}.
\label{coherence}
\eeq
Figure \ref{phiphi}(c) shows that the coherence decays earlier at a higher temperature, which indicates that thermal fluctuations destroy the coherence. 
The asymptotic values of the coherence factor are dependent on the damping, as illustrated in Fig. \ref{coherenceasym}. 
The dependence on $\Omega_{\rc}$ reflects the dependence of the variance as in Eq. \eqref{coherence}. 
One can see that the decay of coherence at a higher temperature gets gradual with larger damping. 
This behavior can be interpreted that the large friction suppresses the deviation of the coherence due to thermal fluctuations, while the coherence would be destroyed by thermal fluctuations with small friction. 
However, the figure indicates that, even in the underdamped limit $\Omega_{\rc}/\Omega_{\J}\to 0$, the coherence keeps finite as $\left\langle\cos{\phi(\infty)}\right\rangle\to\exp\left[-e\Omega_{\J}/I_{\mathrm{c}}\right]$ at $T=0$. 
This result implies that a supercurrent flows even in the underdamped limit at $T=0$ contrary to the picture of Schmid-Bulgadaev transition, which claims that the junction is insulating with a resistance above a critical resistance. 
This consequence is consistent with the recent work in Ref. \cite{murani} on the absence of the Schmid-Bulgadaev transition.

\begin{figure}[t]
\centering
\includegraphics[width=80mm]{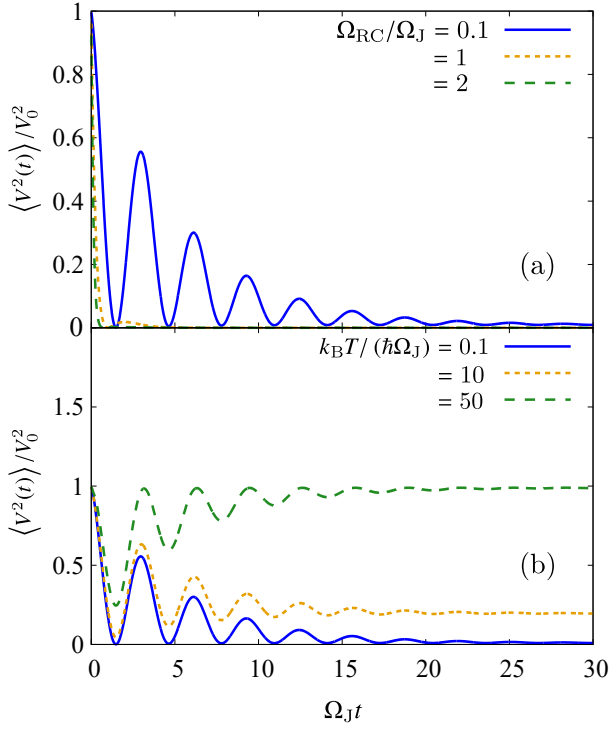}
\caption{Autocorrelation function of the voltage $\left\langle V^{2}(t)\right\rangle$. The upper panel (a) displays the results for $k_{\mathrm{B}}T/\left(\hbar\Omega_{\J}\right)=0.1$. The blue solid, orange dotted, and green dashed lines respectively stand for the results for $\Omega_{\rc}/\Omega_{\J}=0.1$ (underdamped), $\Omega_{\rc}/\Omega_{\J}=1$ (critically damped), and $\Omega_{\rc}/\Omega_{\J}=2$ (overdamped). 
The lower panel (b) displays the results for $\Omega_{\rc}/\Omega_{\J}=0.1$. The blue solid, orange dotted, and green dashed lines respectively stand for the results for $k_{\mathrm{B}}T/\left(\hbar\Omega_{\J}\right)=0.1, 10, 50$. We set $2e\Omega_{\J}/I_{\mathrm{c}}=10^{-2}$ and $V_{0}=\hbar\Omega_{\J}/(2e)$.}
\label{VV}
\end{figure}

In a similar manner, one can obtain the dynamics of the voltage in the absence of noise as
\beq
\bal
V_{0}(t)&=\frac{\Phi_{0}}{2\pi}\dot{\phi}_{0}(t) \\
&=V_{0}\frac{e^{-\Omega_{\rc}t}}{\sqrt{1-\Omega_{\rc}^{2}/\Omega_{\J}^{2}}}
\Bigg[-\frac{\Omega_{\rc}}{\Omega_{\J}}\sin{\left(\sqrt{\Omega_{\J}^{2}-\Omega_{\rc}^{2}}t\right)} \\
&+\sqrt{1-\frac{\Omega_{\rc}^{2}}{\Omega_{\J}^{2}}}\cos{\left(\sqrt{\Omega_{\J}^{2}-\Omega_{\rc}^{2}}t\right)}\Bigg] ,
\eal
\eeq
and the two-point correlation of the voltage as
\beq
\bal
&\left\langle V(t) V(t')\right\rangle
=V_{0}(t)V_{0}(t') \\
&+\left(\frac{\Phi_{0}}{2\pi}\right)^{2}\frac{2}{R}\int^{\infty}_{-\infty}\frac{d\omega}{2\pi}\hbar\omega\coth{\left(\frac{\hbar\omega}{2k_{\mathrm{B}}T}\right)}e^{i\omega(t-t')} \\
&\times\left[\partial_{t}G_{t}(\omega)+i\omega G_{t}(\omega)\right]\left[\partial_{t'}G_{t'}(-\omega)-i\omega G_{t'}(-\omega)\right] .
\eal
\label{VVq}
\eeq
Figure \ref{VV}(a) illustrates the dynamics of the autocorrelation function of the voltage $\langle V^{2}(t)\rangle$ for $k_{\mathrm{B}}T=0.1\hbar\Omega_{\J}$ with different damping constant. 
Similar to that of the phase in Fig. \ref{phiphi}(a), one can see that the correlation of the voltage is strongly suppressed as one increases $\Omega_{\rc}/\Omega_{\J}$. 
The asymptotic correlation in the long-time limit is also dependent on $\Omega_{\rc}$ as in Eq. \eqref{VVqa} as a consequence of the quantum fluctuations similar to the phase correlation. 
As well as the phase correlation, we expect that this dependence on the damping constant can also be experimentally observed. 
The difference from the phase correlation is that, even in the long-time limit, the integral in Eq. \eqref{VVq} does not converge in general [see Eq. \eqref{VVqa} in Appendix \ref{AppC}]. 
The voltage correlation in Eq. \eqref{VVq} indeed converges only in the classical limit with any damping, or in the underdamped limit in any temperature regime.
In the underdamped limit, the asymptotic value converges to $2e\Omega_{\J}/I_{\mathrm{c}}\cdot V_{0}^{2}\coth{\left[\hbar\Omega_{\J}/\left(2k_{\mathrm{B}}T\right)\right]}$, which recovers the classical limit in Eq. \eqref{VVcla}.

Figure \ref{VV}(b) shows the results with different temperature for $\Omega_{\rc}=0.1\Omega_{\J}$. 
As well as the phase correlation in Fig. \ref{phiphi}(b), Fig. \ref{VV}(b) indicates that thermal fluctuations enhance the correlation of voltage in a long time.

\subsection{Effects of external current}

\begin{figure}[t]
\centering
\includegraphics[width=80mm]{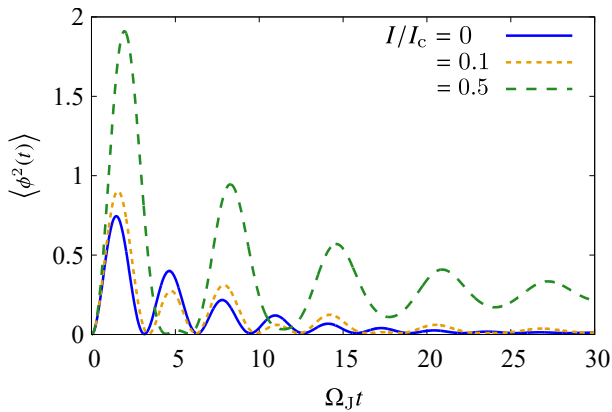}
\caption{Autocorrelation function $\left\langle\phi^{2}(t)\right\rangle$ for $\Omega_{\rc}/\Omega_{\J}=0.1$ and $k_{\mathrm{B}}T/\left(\hbar\Omega_{\J}\right)=0.1$. The blue solid, orange dotted, and green dashed lines respectively stand for the results for $I/I_{\mathrm{c}}=0, 0.1, 0.5$. We set $2e\Omega_{\J}/I_{\mathrm{c}}=10^{-2}$ and $V_{0}=\hbar\Omega_{\J}/(2e)$.}
\label{phiphiIext}
\end{figure}

In the presence of the external current $I$, instead of Eq. \eqref{EoMphi0line}, we solve
\beq
\ddot{\phi}_{0}+2\Omega_{\rc}\dot{\phi}_{0}+\Omega_{\J}^{2}\phi_{0}=\Omega_{\J}^{2}\frac{I}{I_{\mathrm{c}}},
\label{EoMphi0lineext}
\eeq
and the resulting deterministic part of the phase $\phi_{0}$ is given by
\beq
\bal
&\phi_{0}(t)
=\frac{\Omega_{\J}}{\sqrt{\Omega_{\J}^{2}-\Omega_{\rc}^{2}}} \\
&\times\Bigg[\left(\frac{2\pi V_{0}}{\Phi_{0}\Omega_{\J}}-\frac{\Omega_{\rc}}{\Omega_{\J}}\frac{I}{I_{\mathrm{c}}}\right)e^{-\Omega_{\rc}t}\sin{\left(\sqrt{\Omega_{\J}^{2}-\Omega_{\rc}^{2}}t\right)} \\
&+\sqrt{1-\left(\frac{\Omega_{\rc}}{\Omega_{\J}}\right)^{2}}\frac{I}{I_{\mathrm{c}}}
\left[1-e^{-\Omega_{\rc}t}\cos{\left(\sqrt{\Omega_{\J}^{2}-\Omega_{\rc}^{2}}t\right)}\right]\Bigg],
\eal
\eeq
which recovers Eq. \eqref{phi0I0} in $I=0$. 
Since we are working within the linear regime, the transition to the running state due to the tilted washboard potential is absent in this case and the external current just shifts and intensifies the deterministic part of the phase instead, as shown in Fig. \ref{phiphiIext}. 
With a larger external current, the magnitude of phase correlation is more intensified and shifted.
In order to justify the linear approximation we employed in Sec. \ref{SecIIA}, it is required to use the relatively small external current.
Under the original washboard potential in Eq. \eqref{washboard}, with a large external current, the potential has no extremum and the phase flows away from the initial phase as illustrated in Fig. \ref{phiwash}(b). 
This running state strongly enhances the deterministic part of the phase correlation $\left\langle\phi_{0}^{2}(t)\right\rangle$ in the long-time regime, which would make the noise contribution negligible. 
With a sufficiently small current, on the other hand, the phase should oscillate around an extremum of the potential as shown in Fig. \ref{phiwash}(a), and the linear approximation is expected to well describe the dynamics.
Hence, in Fig. \ref{phiphiIext}, we used three relatively small values of external current $I/I_{\mathrm{c}}=0, 0.1, 0.5$.  
In $t\to\infty$, the correlation function asymptotically approaches the nonzero asymptotic value of the deterministic part $\phi^{2}_{0}(t\to\infty)\to I/I_{\mathrm{c}}$ plus that of the variance.

\section{Conclusion}

We have investigated the dynamics of correlations in a RCSJ junction subject to an external current and the current noise generated by a resistor including thermal and quantum fluctuations within the linear regime. The external current affects the deterministic contribution to phase and voltage. It intensifies the phase correlation and makes the oscillation out of phase within the linear regime. We observed the relaxation of correlation functions to finite values due to the current noise with various damping and temperature regime. Due to the zero-point fluctuations in the current noise, we employed the superconducting gap as a cutoff in the variance of phase as well as voltage, which results in suppression of the correlations in the small resistance regime. We also found that the correlations are enhanced by thermal fluctuations at a higher temperature compared to the low-temperature region in which quantum fluctuations are dominant. It results in an earlier decay of coherence at a higher temperature. 
Prominently, we found that the asymptotic correlations in the long-time limit depend on the damping constant or resistance as the emergence of quantum effects, which originate from the quantum fluctuations in the current noise. 
We expect that this dependence on damping in the long-time limit can be detected experimentally and our work can be tested as the emergence of quantum effects in a Josephson circuit.

\begin{acknowledgments}
K.F. is supported by a Ph.D. fellowship of the Fondazione Cassa di Risparmio di Padova e Rovigo. 
This work is partially supported by the BIRD project ``Time-dependent density functional theory of quantum atomic mixtures'' in the University of Padova.
\end{acknowledgments}
%　\clearpage

\appendix

\section{Classical limit}\label{AppA}

In the classical limit $k_{\mathrm{B}}T/\left(\hbar\Omega_{\J}\right)\to\infty$, quantum fluctuations are absent and the noise includes only thermal fluctuations. The classical current noise satisfies
\bseq
\beq
\left\langle\eta(t)\right\rangle=0, 
\eeq
\beq
\left\langle\eta(t)\eta(0)\right\rangle=\frac{4}{R}k_{\mathrm{B}}T \delta(t).
\label{clnoiseb}
\eeq
\label{clnoise}
\eseq
One refers to Eq. \eqref{clnoiseb} as a classical fluctuation-dissipation relation without quantum fluctuations. This classical noise of Eq. \eqref{clnoise} leads to the following autocorrelation functions as
\beq
\bal
&\left\langle\phi^{2}(t)\right\rangle
=\phi_{0}^{2}(t)+\frac{2e\Omega_{\J}}{I_{\mathrm{c}}}\frac{2k_{\mathrm{B}}T}{\hbar\Omega_{\J}} \\
&\times\Bigg[1-\frac{e^{-2\Omega_{\rc}t}}{1-\Omega_{\rc}^{2}/\Omega_{\J}^{2}}\Bigg[1-\left(\frac{\Omega_{\rc}}{\Omega_{\J}}\right)^{2}\cos{\left(2\sqrt{\Omega_{\J}^{2}-\Omega_{\rc}^{2}}t\right)} \\
&+\frac{\Omega_{\rc}}{\Omega_{\J}}\sqrt{1-\left(\frac{\Omega_{\rc}}{\Omega_{\J}}\right)^{2}}\sin{\left(2\sqrt{\Omega_{\J}^{2}-\Omega_{\rc}^{2}}t\right)}\Bigg]\Bigg] .
\eal
\eeq
That for voltage can also be obtained as
\beq
\bal
&\left\langle V^{2}(t)\right\rangle
=V_{0}^{2}(t)+\left(\frac{\hbar\Omega_{\J}}{2e}\right)^{2}\frac{2e\Omega_{\J}}{I_{\mathrm{c}}}\frac{2k_{\mathrm{B}}T}{\hbar\Omega_{\J}} \\
&\times\Bigg[1-\frac{e^{-2\Omega_{\rc}t}}{1-\Omega_{\rc}^{2}/\Omega_{\J}^{2}}\Bigg[1-\left(\frac{\Omega_{\rc}}{\Omega_{\J}}\right)^{2}\cos{\left(2\sqrt{\Omega_{\J}^{2}-\Omega_{\rc}^{2}}t\right)} \\
&-\frac{\Omega_{\rc}}{\Omega_{\J}}\sqrt{1-\left(\frac{\Omega_{\rc}}{\Omega_{\J}}\right)^{2}}\sin{\left(2\sqrt{\Omega_{\J}^{2}-\Omega_{\rc}^{2}}t\right)}\Bigg]\Bigg] .
\eal
\eeq
Since the noise-free contributions $\phi_{0}(t)$ and $V_{0}(t)$ vanish in $t\to\infty$, the asymptotic values for each of the autocorrelations are
\beq
\left\langle\phi^{2}(t\to\infty)\right\rangle
=\frac{2e\Omega_{\J}}{I_{\mathrm{c}}}\frac{2k_{\mathrm{B}}T}{\hbar\Omega_{\J}},
\label{phiphicla}
\eeq
and
\beq
\left\langle V^{2}(t\to\infty)\right\rangle
=\left(\frac{\hbar\Omega_{\J}}{2e}\right)^{2}\frac{2e\Omega_{\J}}{I_{\mathrm{c}}}\frac{2k_{\mathrm{B}}T}{\hbar\Omega_{\J}},
\label{VVcla}
\eeq
which are proportional to the temperature reflecting the thermal noise in Eqs. \eqref{clnoise}, so that both of them flow to infinity in the classical limit $k_{\mathrm{B}}T/\left(\hbar\Omega_{\J}\right)\to\infty$.

\section{Expression of $G_{t}(\omega)$}\label{AppB}

Here we explicitly write down the expression of $G_{t}(\omega)$ which is necessary to obtain the autocorrelation functions. Using Eq. \eqref{Gt} and Eq. \eqref{Gomega}, one immediately obtains
\beq
\bal
G_{t}(\omega)&=-\frac{\Omega_{\J}^{2}}{2I_{\mathrm{c}}}\frac{1}{\sqrt{\Omega_{\J}^{2}-\Omega_{\rc}^{2}}} \Bigg[\frac{e^{i\left(i\Omega_{\rc}+\sqrt{\Omega_{\J}^{2}-\Omega_{\rc}^{2}}-\omega\right)t}-1}{i\Omega_{\rc}+\sqrt{\Omega_{\J}^{2}-\Omega_{\rc}^{2}}-\omega} \\
&-\frac{e^{i\left(i\Omega_{\rc}-\sqrt{\Omega_{\J}^{2}-\Omega_{\rc}^{2}}-\omega\right)t}-1}{i\Omega_{\rc}-\sqrt{\Omega^{2}_{\J}-\Omega_{\rc}^{2}}-\omega}\Bigg].
\eal
\eeq
As for the correlation of voltage, we need to compute a quantity $\left[\partial_{t}+i\omega\right]G_{t}(\omega)$. It is given by
\beq
\bal
&\left[\partial_{t}+i\omega\right]G_{t}(\omega)
=-\frac{i\Omega_{\J}^{2}}{2I_{\mathrm{c}}}\frac{1}{\sqrt{\Omega_{\J}^{2}-\Omega_{\rc}^{2}}} \\
&\times\Bigg[\frac{i\Omega_{\rc}+\sqrt{\Omega_{\J}^{2}-\Omega_{\rc}^{2}}}{i\Omega_{\rc}+\sqrt{\Omega_{\J}^{2}-\Omega_{\rc}^{2}}-\omega}\left[e^{i\left(i\Omega_{\rc}+\sqrt{\Omega_{\J}^{2}-\Omega_{\rc}^{2}}-\omega\right)t}-1\right] \\
&-\frac{i\Omega_{\rc}-\sqrt{\Omega_{\J}^{2}-\Omega_{\rc}^{2}}}{i\Omega_{\rc}-\sqrt{\Omega_{\J}^{2}-\Omega_{\rc}^{2}}-\omega}\left[e^{i\left(i\Omega_{\rc}-\sqrt{\Omega_{\J}^{2}-\Omega_{\rc}^{2}}-\omega\right)t}-1\right]
\Bigg].
\eal
\eeq

\section{Long-time limit}\label{AppC}

Correlation functions in the long-time limit can also be obtained through Fourier analysis \cite{josephson,fordPLA}. 
Performing the Fourier transformation on Eq. \eqref{EoMdelphiline}, we obtain \cite{fordPLA}
\beq
\bal
\delta\Tilde{\phi}(\omega)\equiv \int^{\infty}_{-\infty}dt\delta\phi(t)e^{-i\omega t}
=\alpha(\omega)\Omega_{\J}^{2}\frac{\Tilde{\eta}(\omega)}{I_{\mathrm{c}}} ,
\eal
\eeq
where $\Tilde{\eta}(\omega)\equiv \int^{\infty}_{-\infty}dt\eta(t)e^{-i\omega t}$ and
\beq
\alpha(\omega)\equiv \frac{1}{-\omega^{2}+2i\Omega_{\rc}\omega+\Omega_{\J}^{2}}.
\eeq
Remarkably, $\alpha(\omega)$ is equivalent to the long-time limit of $G_{t}(\omega)$ in Eq. \eqref{Gomega} as
\beq
\alpha(\omega)=\frac{I_{\mathrm{c}}}{\Omega_{\J}^{2}}G_{t\to\infty}(\omega).
\eeq
Hence the time-independent autocorrelation function with respect to the phase in the long-time limit can be obtained as \cite{fordPLA}
\beq
\left\langle\phi^{2}(t\to\infty)\right\rangle
=\frac{2\Omega_{\J}^{4}}{RI_{\mathrm{c}}^{2}}\int^{\infty}_{-\infty}\frac{d\omega}{2\pi}\hbar\omega\coth{\left(\frac{\hbar\omega}{2k_{\mathrm{B}}T}\right)}\abs{\alpha(\omega)}^{2}.
\label{phiphiqa}
\eeq
In particular, in the underdamped limit $\Omega_{\rc} \ll \Omega_{\J}$, by virtue of the following relation
\beq
\bal
2\Omega_{\rc}\omega\abs{\alpha(\omega)}^{2} 
\to\frac{\pi}{2\Omega_{\J}}\left[\delta(\omega-\Omega_{\J})+\delta(\omega+\Omega_{\J})\right],
\eal
\eeq
one obtains
\beq
\left\langle\phi^{2}(t\to\infty)\right\rangle
\to\frac{2e\Omega_{\J}}{I_{\mathrm{c}}}\coth{\left(\frac{\hbar\Omega_{\J}}{2k_{\mathrm{B}}T}\right)},
\label{phiphiua}
\eeq
which is consistent with the classical limit in Eq. \eqref{phiphicla}.

In a similar manner, one can write the autocorrelation function with respect to the voltage as well. It is given by
\beq
\bal
&\left\langle V^{2}(t\to\infty)\right\rangle \\
&=\left(\frac{\Phi_{0}}{2\pi}\right)^{2}\frac{2\Omega_{\J}^{4}}{RI_{\mathrm{c}}^{2}}\int^{\infty}_{-\infty}\frac{d\omega}{2\pi}\hbar\omega\coth{\left(\frac{\hbar\omega}{2k_{\mathrm{B}}T}\right)}\omega^{2}\abs{\alpha(\omega)}^{2}.
\eal
\label{VVqa}
\eeq
This energy integral is logarithmically divergent, which is distinct from the case of phase correlation in Eq. \eqref{phiphiqa}. In the underdamped limit, it converges and reduces to
\beq
\left\langle V^{2}(t\to\infty)\right\rangle
\to\left(\frac{\hbar\Omega_{\J}}{2e}\right)^{2}\frac{2e\Omega_{\J}}{I_{\mathrm{c}}}\coth{\left(\frac{\hbar\Omega_{\J}}{2k_{\mathrm{B}}T}\right)},
\eeq
which is also consistent with the classical limit in Eq. \eqref{VVcla}.

\end{document}